\documentclass{ws-procs11x85}
\usepackage{multicol,float}
\usepackage{amsmath,mathrsfs,amssymb}
\usepackage{subfigure}
\begin{document}

\title{Identification and Query of Activated Gene Pathways in Disease Progression}

\author{Arvind Rao\footnote{Corresponding author.}, Alfred O. Hero, III }

\address{Machine Learning Department, Computer Sciences Department and Lane Center for Computational Biology, Carnegie Mellon University ,\\
Pittsburgh, PA 15213, USA\\
 $^*$Email: [ukarvind]@cmu.edu}

\address{Electrical Engineering and Computer Science, Bioinformatics, University of Michigan,\\
Ann Arbor, MI 48109, USA\\
Email: [hero]@umich.edu}

%\author{David J. States}
%
%\address{Bioinformatics, Human Genetics, University of Michigan,\\
%Ann Arbor, MI 48109, USA\\
%Email: dstates@umich.edu}
%
%\author{James Douglas Engel}
%
%\address{Cell and Developmental Biology, University of Michigan,\\
%Ann Arbor, MI 48109, USA\\
%Email: engel@umich.edu}

\maketitle
\begin{abstract}
Disease occurs due to aberrant expression of genes and modulation of the biological pathways along which they lie. Inference of activated gene pathways, using gene expression data during disease progression, is an important problem. In this work, we have developed a generalizable framework for the identification of interacting pathways while incorporating biological realism, using functional data analysis and manifold embedding techniques. Additionally, we have also developed a new method to query for the differential co-ordinated activity of any desired pathway during disease progression. The methods developed in this work can be generalized to any conditions of interest. %the method explores an hitherto unexamined way of querying for co-ordinated pathway activity .
\end{abstract}

\keywords{Functional Data Analysis (FDA), Gene ontology, immune response, Laplacian Eigenmaps, Mantel correlation, logDet divergence, functional genomics, heterogeneous data integration.}

\bodymatter
\begin{multicols}{2}

\section{INTRODUCTION}

A gene is a segment of the genome (DNA) that codes for protein. Proteins are the biological catalysts of function and mediate the kinetics and timing of any biological process. The body is composed of several tissues and each tissue has its own individual lineage of cells. The set of proteins that are active in each cell are different depending on cell context and the underlying function of the cell. Since proteins are encoded by the gene(s), it is believed that biological processes are orchestrated by a precise spatio-temporal expression of genes. Gene activities in the cell are organized along pathways, which might either have signaling activity or regulatory role in determining cell behavior. Thus even though there is a heterogeneous population of cells in the body, each set of cells has a precise time and location of gene/pathway expression corresponding to required protein activity.

Systemic disease can occur due to the mis-expression of genes in various tissues [as  an example, the \textit{Gata3} gene is mutated in HDR (hypoparathyroidism, deafness, renal dysplasia)]. A lot of studies have studied the link between the genome and the phenome - i.e. between gene expression and physical characteristic. Disease (and its symptoms) is a physical characteristic which arises from mis-expression of the underlying genome. Also, very rarely is disease due to the aberrant expression of a single gene - they mostly arise due to mis-expression of several genes at once (i.e. gene sets). Identifying these set of aberrant genes (or pathways) is an important problem because of the immense therapeutic potential. There is an ongoing effort to find inhibitors that can target disease-implicated pathways. For example, two well-known drugs Gleevec and Tarceva target receptor tyrosine kinase signaling pathways that have a reported role in cancer [\refcite{RTK_book}].% Role in cancer [RTKs for cancer: Gleevec, Tarceva]

In this work, we develop methods to identify pathways (Part $I$) that have a role in disease progression. Specifically, we ask which pathways are potentially modulated during onset and evolution of immune response to infection. Additionally, we present a framework (Part $II$) that can query the differential activity of `any' pathway between normal and diseased states, thereby allowing for the principled selection of pathway inhibitors to modulate and control disease. Using time series expression profiles of gene expression, we use functional data analysis (FDA) to process, analyze and cluster the data into possible pathway components. In addition, we use a manifold embedding technique to improve on these results and extend this for generalized pathway querying. \baselineskip=13.5pt

\section{OUTLINE}\label{gene_nets}
\label{sec:format}

This paper is organized as follows. To interpret pathway activity during immune response, section \ref{data_processing} deals with the examination of the gene expression data during pathogen infection, and its representation in terms of B-spline basis functions. Section \ref{fPCA} uses principal component analysis on the functional data (fPCA) to discover modes of variation in the data. This is followed by clustering genes in fPCA space to find genes whose interaction is putatively associated with infection. We find that the results are not directly relevant biologically and hence, section \ref{biological_realism} develops methods to improve the context of the clustering results to obtain more meaningful results. This new framework enables the solution of another hitherto unexamined problem - querying \textit{any} arbitrary pathway for co-ordinated differential activity between case and condition (section \ref{part2}). As an example, we examine the activity of the Toll-like receptor (TLR) pathway for the pathogen infection data set, and demonstrate the general utility of this approach in such problems. Section \ref{conclusions} concludes the paper.

\section{DATA EXTRACTION AND PRE-PROCESSING}\label{data_processing}
One of the most common processes involving gene-pathway modulation is the systemic response of the immune system to an invading pathogen. With the advent of whole genome microarrays that can assay the activity of genes over a time course, expression profiling of genes during the innate and adaptive immune response has been actively pursued for the identification of genes that can be used for diagnostic or therapeutic purposes. For this study, in order to find pathways implicated in the immune response to pathogen infection, we use functional expression data gathered by the Young group [\refcite{Young_data}]. This data profiles the various gene activation programs initiated in macrophage cells on exposure to various pathogens such as tuberculosis, e.coli and staphylococcus aureus. There is also a set of control treatments in which latex beads coated without bacteria are presented to the macrophage cell population. In this study, we consider the differential pathway activity between tuberculosis and control conditions. The methods developed in our approach are however, more general and can be applied for any number of interesting conditions.

The dataset contains expression values at $8$ time points for $168$ unique macrophage genes. These correspond to gene expression profiling $0.5,1,3,6,12,16,18,24$ hours after exposure with the pathogen (or control). Since macrophages exhibit an early innate as well as late adaptive immune response, it is interesting to examine which genes are expressed in a certain phase. %The innate response corresponds to the first line of defense to a invading pathogen and involves secretion of several interleukins and cytokines. Once the immune system identifies the

%The raw data is displayed in Fig. \ref{fig:sub_unsmooth_a} and Fig. \ref{fig:sub_unsmooth_b}.
To use the functional data approach on the raw data, we start by representing the functional data via B-spline basis functions. This choice of basis functions is primarily governed by the lack of inherent periodicity in immune response over the first $24$ hours, as well as the possibility of local structure in the time series that are are relevant to analysis. Using the ``FDA" package in R, we create B-spline basis functions ($B_k(t_j)$) of order $3$ and $J=46$ internal knots over the interval $[0.5,24]$. Under this representation, we have,\\

$x_i(t_j)=\sum_{k=1}^K c_k B_k(t_j)$ for $i=1,2,\ldots, n$ with $n=168$.\\% and $K = 48 = m+L-1; m=3; L=47$.\\
%
%\begin{figure}[H]
%\centering
%\subfigure[Unsmoothed Control Data.] % caption for subfigure a
%{
%    \label{fig:sub_unsmooth_a}
%    \includegraphics[width=6cm]{unsmoothed_latex_data.ps}
%}
%\hspace{1cm}
%\subfigure[Unsmoothed Tuberculosis Data.] % caption for subfigure b
%{
%    \label{fig:sub_unsmooth_b}
%    \includegraphics[width=6cm]{unsmoothed_tub_data.ps}
%}
%\caption{Unsmoothed functional data for immune response under tuberculosis and control.}
%\label{fig:sub_unsmooth} % caption for the whole figure
%\end{figure}

\begin{figure}[H]
\centering
\subfigure[Smoothed Control time-series Data.] % caption for subfigure a
{
    \label{fig:sub_smooth_a}
    \includegraphics[scale=0.3]{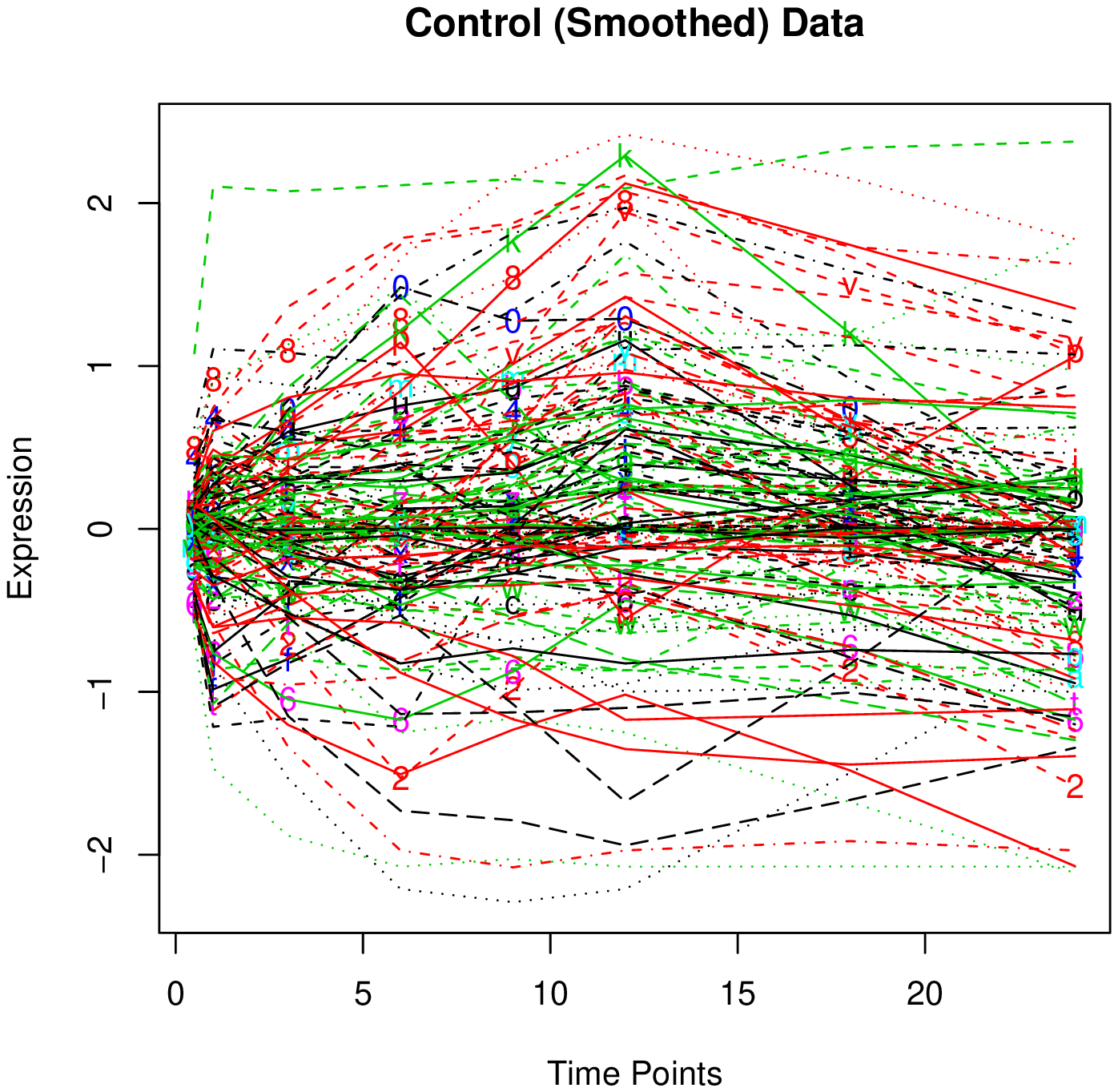}
}
\hspace{1cm}
\subfigure[Smoothed Tuberculosis time-series Data.] % caption for subfigure b
{
    \label{fig:sub_smooth_b}
    \includegraphics[scale=0.3]{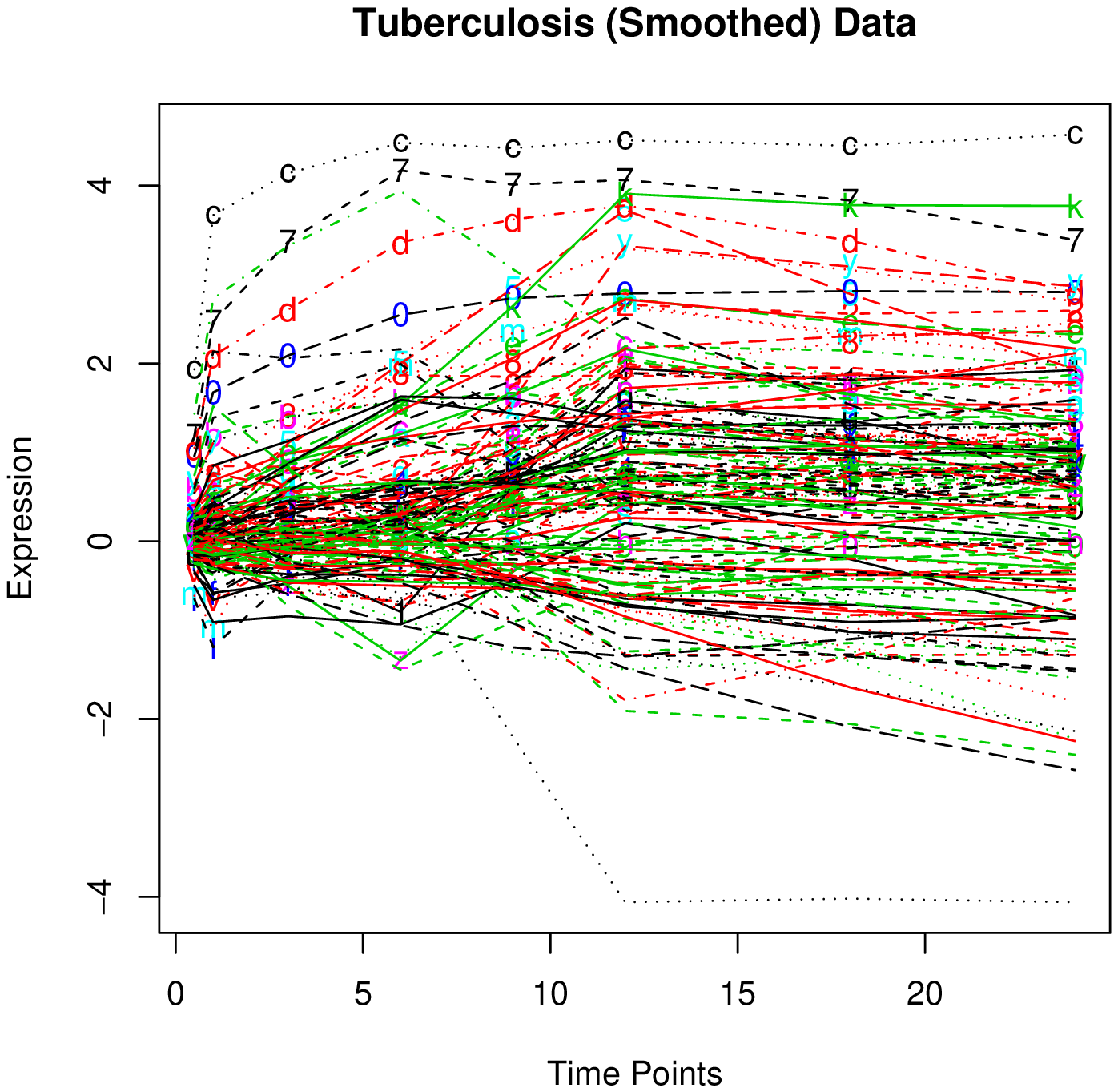}
}
\caption{Smoothed functional data for immune response under tuberculosis and control. The x-axis denotes the time points in hours.}
\label{fig:sub_smooth} % caption for the whole figure
\end{figure}

Furthermore, a smoothing operation is implemented on the data (using generalized cross-validation), with $\lambda=0.001$. The plots of the functional data after smoothing are displayed in Fig. \ref{fig:sub_smooth_a} and Fig. \ref{fig:sub_smooth_b} respectively.

\section{FUNCTIONAL PRINCIPAL COMPONENT ANALYSIS  (fPCA)}\label{fPCA}

%brief overview of fPCA analysis, and varimax interp.
Functional PCA (fPCA) aims to find a  solution to the eigenvalue problem [\refcite{ramsay_1997}, \refcite{ramsay_2002}] : $C \mathbf{\phi} \textbf{b} = \lambda \textbf{b}$,
where $C= [\sum_{i=1}^n c_{i,k} c_{i,l}/n]$,
$\phi= [\langle B_k, B_m\rangle]$ and $\mathbf{b} = (b_1,b_2,\ldots, b_k)$.
The $j^{th}$ principal component eigenvector $\mathbf{b}_j$ of $C\phi$
leads to an estimate $\epsilon_j = [B_1,B_2,\ldots, B_K]^T \mathbf{b}_j$ of the eigenfunction. With this, the $j^{th}$ principal component score is given by $\alpha_{i,j}= \langle x_i,\epsilon_j \rangle$. The set of scores ($[\alpha_{i,1}, \alpha_{i,2},\ldots,\alpha_{i,p}] \in \mathbb{R}^p; i = 1,2,\ldots, 168$) can then be used for clustering [\refcite{Morris_fda}].

\begin{figure}[H]%give PCs for control
\centering
\subfigure[] % caption for subfigure a
{
    \label{fig:sub_tub_pc1}
    \includegraphics[scale=0.26]{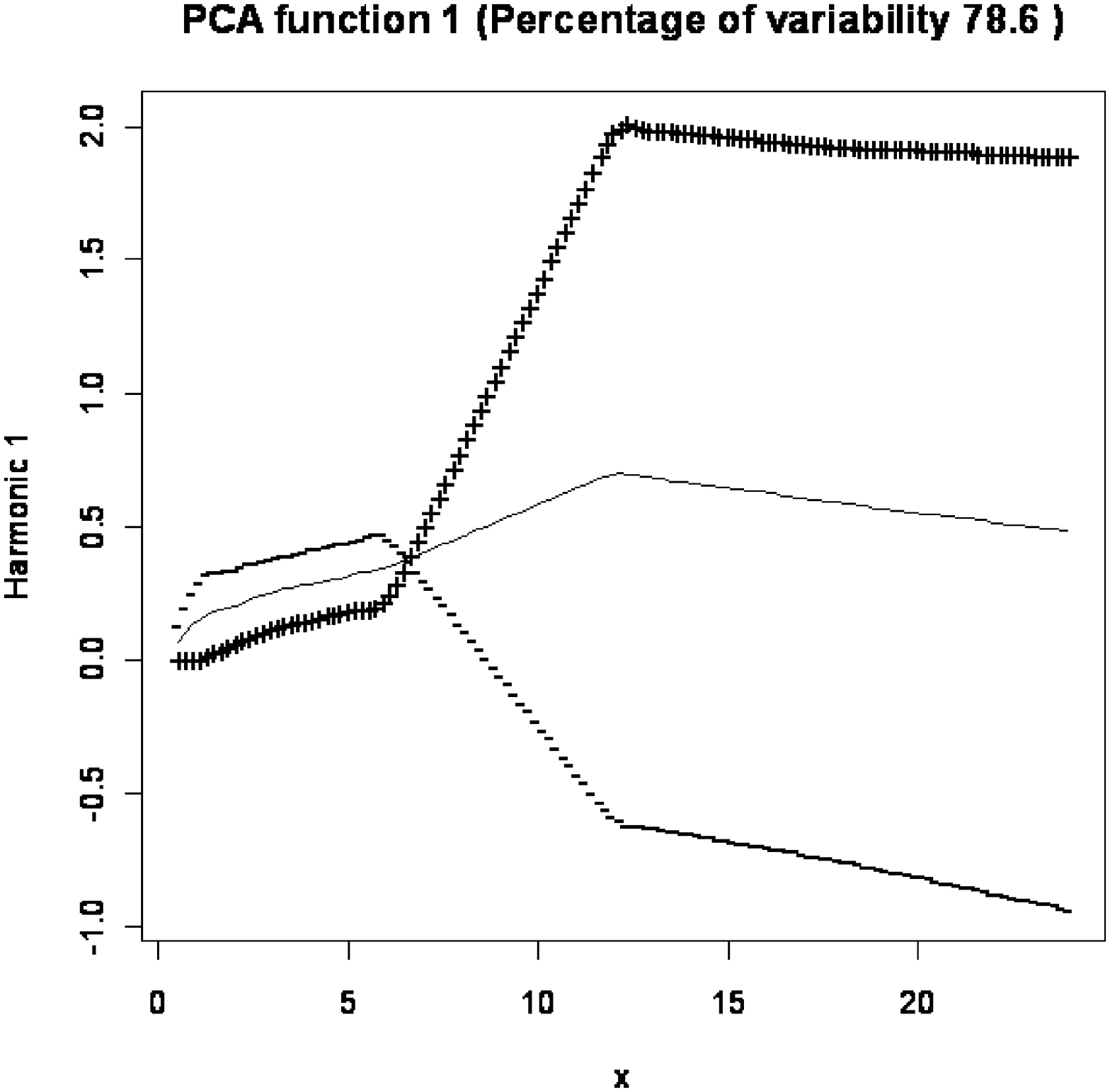}
}
\hspace{1cm}
\subfigure[] % caption for subfigure b
{
    \label{fig:sub_tub_pc2}
    \includegraphics[scale=0.26]{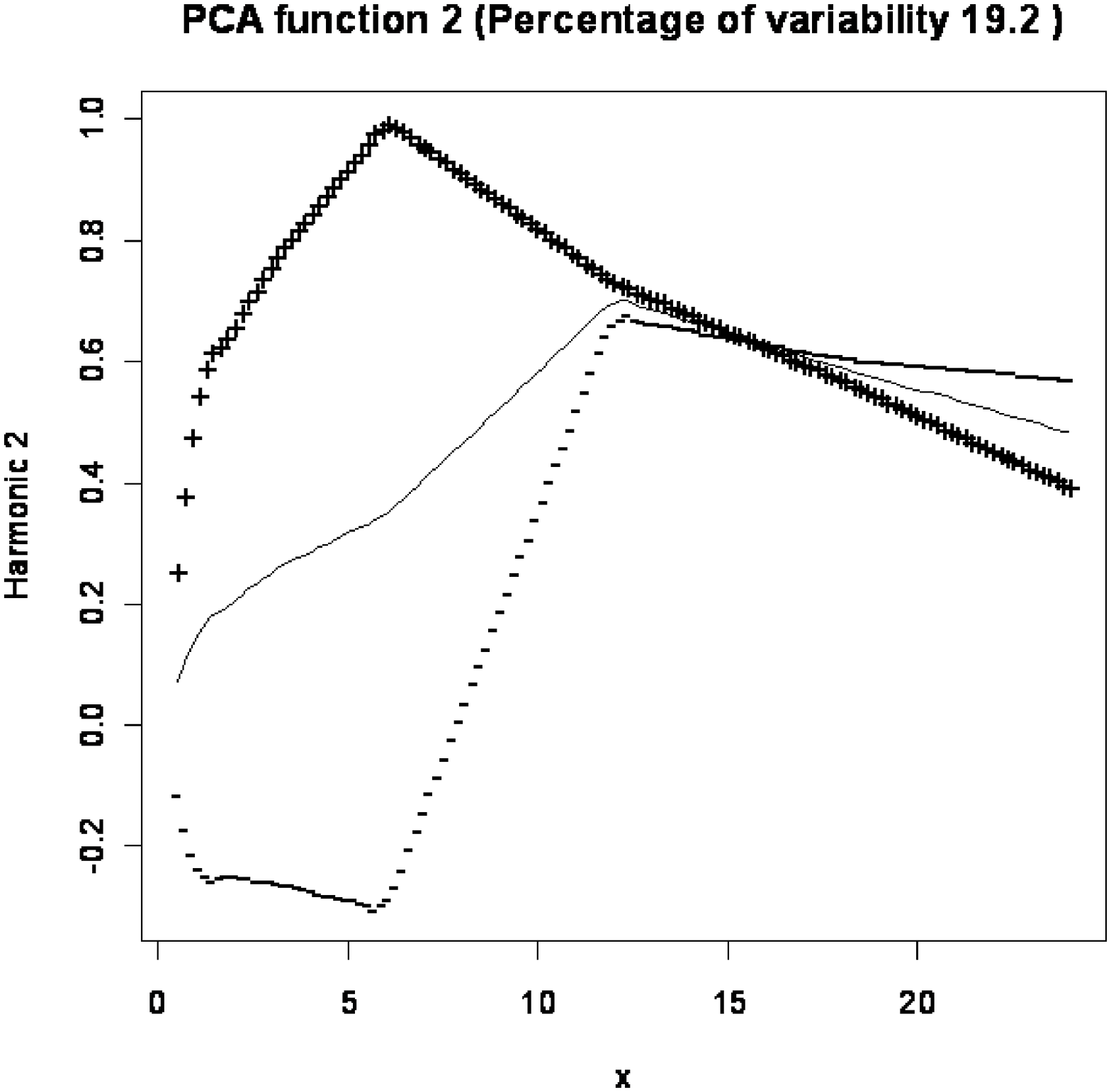}
}
\caption{First and Second Functional Principal Components for Tuberculosis data. X-axis corresponds to assay time points, as in the previous plots. We note that the solid line in each case corresponds to the mean function, the $(+)$ line corresponds to the mean function added to a multiple of the harmonic and the $(-)$ line corresponds to a subtraction of the harmonic function.}
\label{fig:sub_tub_pc} % caption for the whole figure
\end{figure}

To understand the modes underlying disease onset and its response by the immune system, a fPCA analysis of the data was done. The first and second principal components of the tuberculosis functional data are displayed in Fig. \ref{fig:sub_tub_pc} (a) and (b) respectively.  These harmonics correspond to the components after varimax rotation to aid interpretability (\refcite{ramsay_1997},\refcite{ramsay_2002}). The harmonic plots indicate two distinct behaviors and are indicative of typical immune response.

Some interesting insights emerge from the plots of Fig. \ref{fig:sub_tub_pc}. The first harmonic corresponds (roughly) to the `late' variation in gene expression (Fig. \ref{fig:sub_tub_pc1}, accounting for $\sim78\%$ variation) whereas the second principal component corresponds to the `early' variation (Fig. \ref{fig:sub_tub_pc2}, accounting for $\sim20\%$ variation). This is extremely meaningful because the principal components correspond to a drastic change in adaptive immune response and is strongly associated with biological response to pathogen infection. This is also known from biological literature [\refcite{Young_data}, \refcite{Huang_Hacohen}].

%Basic fPCA and its interpretation, for control and tuberculosis

\begin{figure}[H]%give PCs for control
\centering
\subfigure[Scores of functional data along functional principal components, PC1 and PC2] % caption for subfigure a
{
    \label{fig:score_fpca}
    \includegraphics[width=6cm]{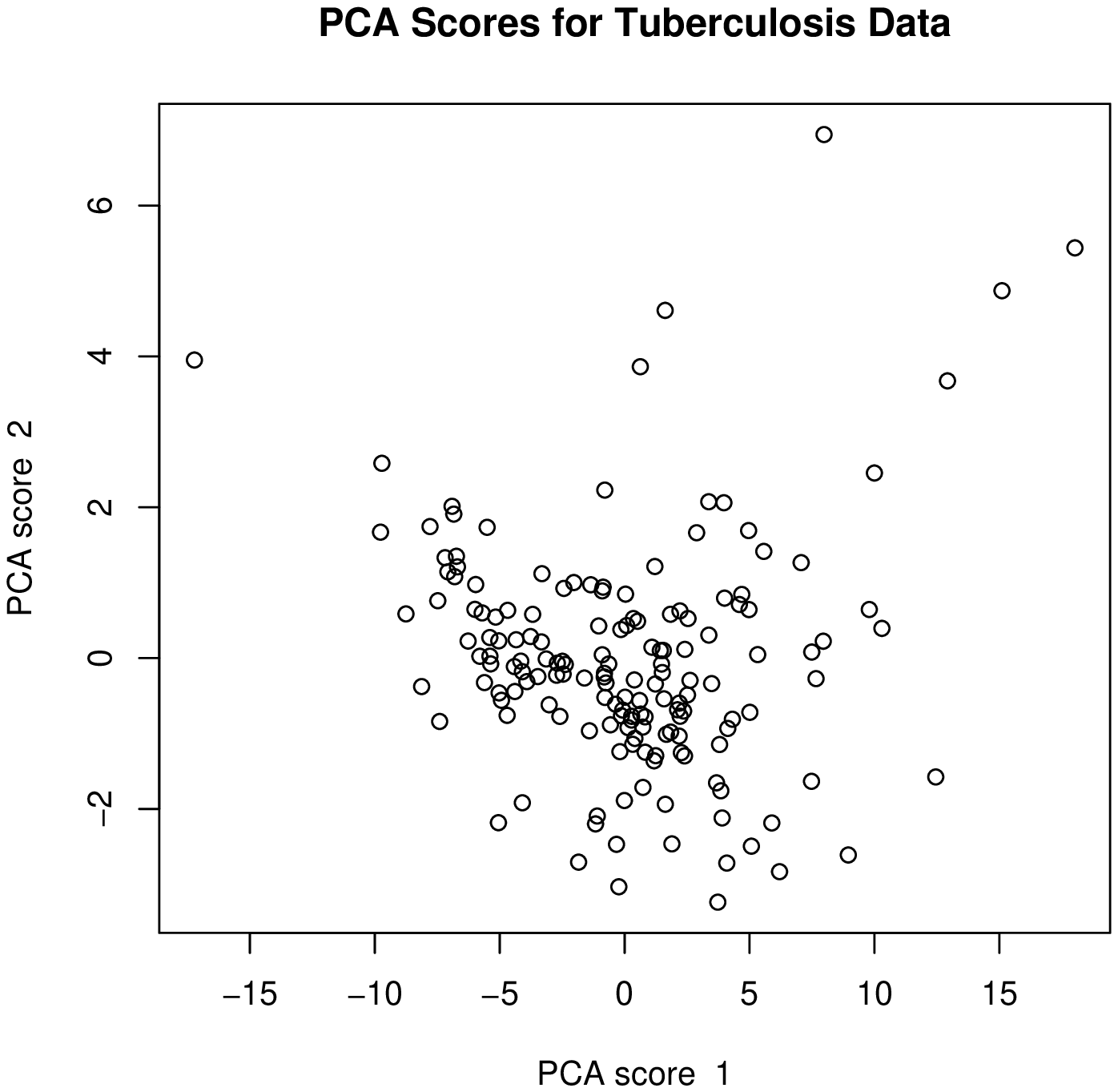}
}
\hspace{1cm}
\subfigure[BIC plot during model-based clustering] % caption for subfigure b
{
    %\label{fig:sub:b}
    \includegraphics[width=6cm]{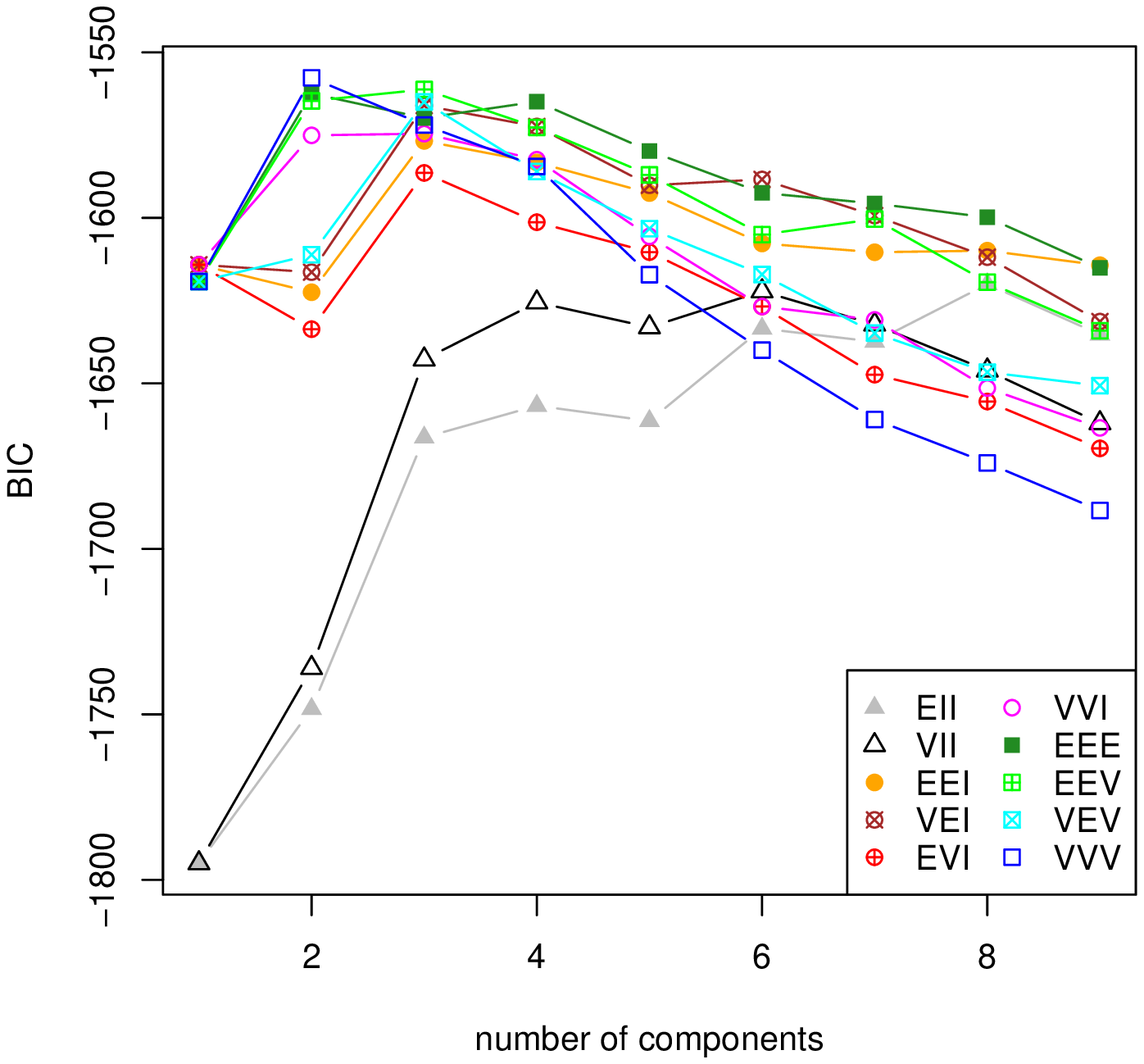}
}
\caption{Embedding of data onto fPCA space and BIC plot to find optimal cluster number.}
\label{fig:tub_pc_cluster} % caption for the whole figure
\end{figure}

%\vspace{-1cm}
The scores of the functional tuberculosis gene data along these first two principal components is shown in Fig. \ref{fig:score_fpca}.

\section{MODEL-BASED CLUSTERING}\label{mclust}
Having found scores of each of the genes in fPCA space, our goal is to now group (cluster) genes with similar temporal profiles. In this section, we derive the parameter update equations for a \textit{Mixture of Gaussian} clustering paradigm [\refcite{MClust_TR}, \refcite{GMM_PAMI}].

We consider the group of gene expression profiles $\mathscr{G} = \{\boldsymbol{g^{(1)}},\boldsymbol{g^{(2)}},\ldots, \boldsymbol{g^{(n)}} \}$, all of which share a common dynamic. Consider gene profile $i$,
$\boldsymbol{g^{(i)}} = [\alpha_{i,1},\alpha_{i,2},\ldots,\alpha_{i,J}]$, a %[g_{1}^{(i)},g_{2}^{(i)},\ldots,g_{T}^{(i)}]^{T}$, a
$J$-dimensional random vector (here $J=2$) which follows a $k$-component finite mixture distribution described by:
\begin{align}
p(\textbf{g}|\boldsymbol{\theta}) &= \sum_{m=1}^{k} \alpha_m
p(\textbf{g}|\boldsymbol{\phi_{m}})
%have to make \theta bold
\end{align}
where $\alpha_1,\ldots,\alpha_k$ are the mixing probabilities, each $\phi_m$ is the set of parameters defining the $m^{th}$ component, and $\boldsymbol{\theta} \equiv
\{\phi_1,\ldots,\phi_k,\alpha_1,\ldots,\alpha_k\}$ is the set of complete parameters needed to specify the mixture. We have,

\begin{align}
\alpha_m\geq0, m=1,\ldots,k, \quad and \quad
\sum_{m=1}^{k}\alpha_m = 1
\end{align}

\vspace{-1cm}

For a set of $n$ independently and identically distributed samples,
\begin{align}
\mathscr{G} &=
\{\boldsymbol{g^{(1)}},\boldsymbol{g^{(2)}},\ldots,\boldsymbol{g^{(n)}}\},
\end{align}
 the log-likelihood of a $k$-component mixture is given by:
 \begin{align}
\log p(\mathscr{G}|\boldsymbol{\theta}) &= \log
\prod_{i=1}^{n} p(\boldsymbol{g^{(i)}}|\boldsymbol{\theta}) \\
&\quad = \sum_{i=1}^{n}log\sum_{m=1}^{k}\alpha_m
p(\boldsymbol{g^{(i)}}|\phi_m)
\end{align}

\begin{itemize}
\item
Treat the labels, $\mathscr{Z} = \{\boldsymbol{z^{(1)}},\ldots,\boldsymbol{z^{(n)}}\}$, associated with the $n$ samples - as missing data. Each label is a binary vector $\boldsymbol{z^{(i)}} = [z_1^{(i)},\ldots,z_k^{(i)}]$, where $z_m^{(i)} =1$ and
$z_p^{(i)}=0$, for $p \ne m$ indicates that sample $\boldsymbol{g^{(i)}}$ was
produced by the $m^{th}$ component.
\end{itemize}

In this setting,  the \textbf{Expectation Maximization} algorithm can be used
to derive the cluster parameter $(\boldsymbol{\theta})$ update equations.

In the \emph{E step} of the \emph{EM algorithm}, the function
 $Q(\boldsymbol{\theta},\boldsymbol{\hat{\theta}}(t)) \equiv
 E[\textrm{log} p(\mathscr{G,Z}|\boldsymbol{\theta})|\mathscr{G},\hat{\boldsymbol{\theta}}(t)], $is
 computed. This yields,
 \begin{align}
 w_m^{(i)} \equiv E[z_m^{(i)}|\mathscr{G},\hat{\theta}_t] &= \frac{\hat{\alpha}_m (t) p(\boldsymbol{g^{(i)}}|\boldsymbol{\hat{\theta}_m} (t))}{\sum_{j=1}^{k}\hat{\alpha}_j (t) p(\boldsymbol{g^{(i)}}|\boldsymbol{\hat{\theta}_j} (t))}
 \end{align}
 where $w_m^{(i)}$ is the posterior probability of the event
 $z_m^{(i)} = 1$, on observing $g_m^{(i)}$.
 \begin{align}
%\intertext{The estimate of the number of components ($k$) is chosen
%using a Minimum Message Length (MML) criterion [2]. The MML
%criterion borrows from algorithmic information theory and serves to
%select models of lowest complexity to explain the data. As can be
%seen below, this complexity has two components - the first encodes
%the observed data as a function of the model and the second encodes
%the model itself. Hence, the MML criterion in our setup becomes, :}
%\hat{k}_{MML} &= \textrm{argmin}_k \{-\textrm{log}
%p(\mathscr{G}|\boldsymbol{\hat{\theta}}(k)) + \frac{k(N_p+1)}{2}
%\textrm{log} n \},
%\end{align}
% $N_p$ is number of parameters
%per component in the $k$ component mixture, given the number of
%clusters $k_{min} \le k \le k_{max}$. \\%, given $k_{min}$ and $k_{max}.$}
%
%
%Alternatively, the bayesian information criterion is used, $\hat{k}_{BIC} = argmax_k\{2l_M(y,\hat{\theta})-m_M log(n)\}$\\
%here, the $l_M(y,\hat{\theta})$ term represents the
\intertext{The estimate of the number of components ($k$) is chosen
using a bayesian information criterion (BIC) criterion [\refcite{MClust_TR}, \refcite{GMM_PAMI}]. The BIC
criterion borrows from information theory and serves to
select models of lowest complexity to explain the data. As can be
seen below, this complexity has two components - the first encodes
the observed data as a function of the model and the second encodes
the model itself. Hence, the BIC criterion in our setup becomes, :}
\hat{k}_{BIC} &= \textrm{argmax}_k \{2\textrm{log}
p(\mathscr{G}|\boldsymbol{\hat{\theta}}(k)) - \frac{k(N_p+1)}{2}
\textrm{log} n \},
\end{align}
 $N_p$ is number of parameters
per component in the $k$ component mixture, given the number of
clusters $k_{min} \le k \le k_{max}$. $n$ is the total number of observations.\\%, given $k_{min}$ and $k_{max}.$}

\vspace{-0.8cm}
\begin{gather}
\intertext{In the \emph{M step}:
For $m=0,1,\ldots,k,$
$\hat{\theta}_m(t+1) = \arg \max_{\phi_m}
Q(\boldsymbol{\theta},\hat{\boldsymbol{\theta}}(t))$, for $m: \hat{\alpha}_m(t+1) >
0$, the elements $\hat{\phi}$'s of the parameter vector estimate $\boldsymbol{\hat{\theta}}$ are typically not closed form and depend on the specific parametrization of the densities in in the mixture, i.e. $p(\boldsymbol{g^{(i)}}|\boldsymbol{\phi_m)}$.
If $p(\textbf{g}^{(i)}|\phi_m)$ belongs to the Gaussian density
$\mathscr{N}(\boldsymbol{\mu}_{m},\boldsymbol{\Sigma}_{m})$ class, we have,
$\phi = (\boldsymbol{\mu},\boldsymbol{\Sigma})$ and EM updates yield \cite{GMM_PAMI},}
\hat{\alpha}_m(t+1) = \frac{\sum_{i=1}^{n}w_m^{(i)}}{n} , \\
\boldsymbol{\mu}_m(t+1) =
\frac{\sum_{i=1}^{n}w_m^{(i)}\textbf{g}^{(i)}}{\sum_{i=1}^{n}w_m^{(i)}},
\\
\boldsymbol{\Sigma}_m(t+1) = \frac{\sum_{i=1}^{n}
w_m^{(i)}(\textbf{g}^{(i)}-\mu_m(t+1))(\textbf{g}^{(i)}-\mu_m(t+1))^{T}}{\sum_{i=1}^{n}w_m^{(i)}}
\end{gather}

The equations \ref{mclust}. $6, 8, 9, 10$ are the parameter update equations for each of the $m = 1, \ldots k$ cluster components.

The R package `MClust' can be used to do model based clustering on the genes (represented as fPCA scores). The method enables the selection of the optimal number of clusters via maximizing the bayesian information criterion (BIC). It outputs a plot that displays the BIC for each cluster assignment, where the clusters can have three independent degrees of freedom (shape, volume and orientation). The optimal cluster assignment is chosen from all possibilities of shape, volume, orientation and BIC [\refcite{MClust_TR}].

%\textit{Note}: We note that even though, we consider each gene to be a $T$-dimensional vecotr, the clustering procedure

\section{fPCA CLUSTERING RESULTS}\label{fpca_cluster}
\begin{figure}[H]
\centering
\subfigure[BIC based model clustering.] % caption for subfigure a
{
    \label{fig:fpca_tub_bic}
    \includegraphics[width=6cm]{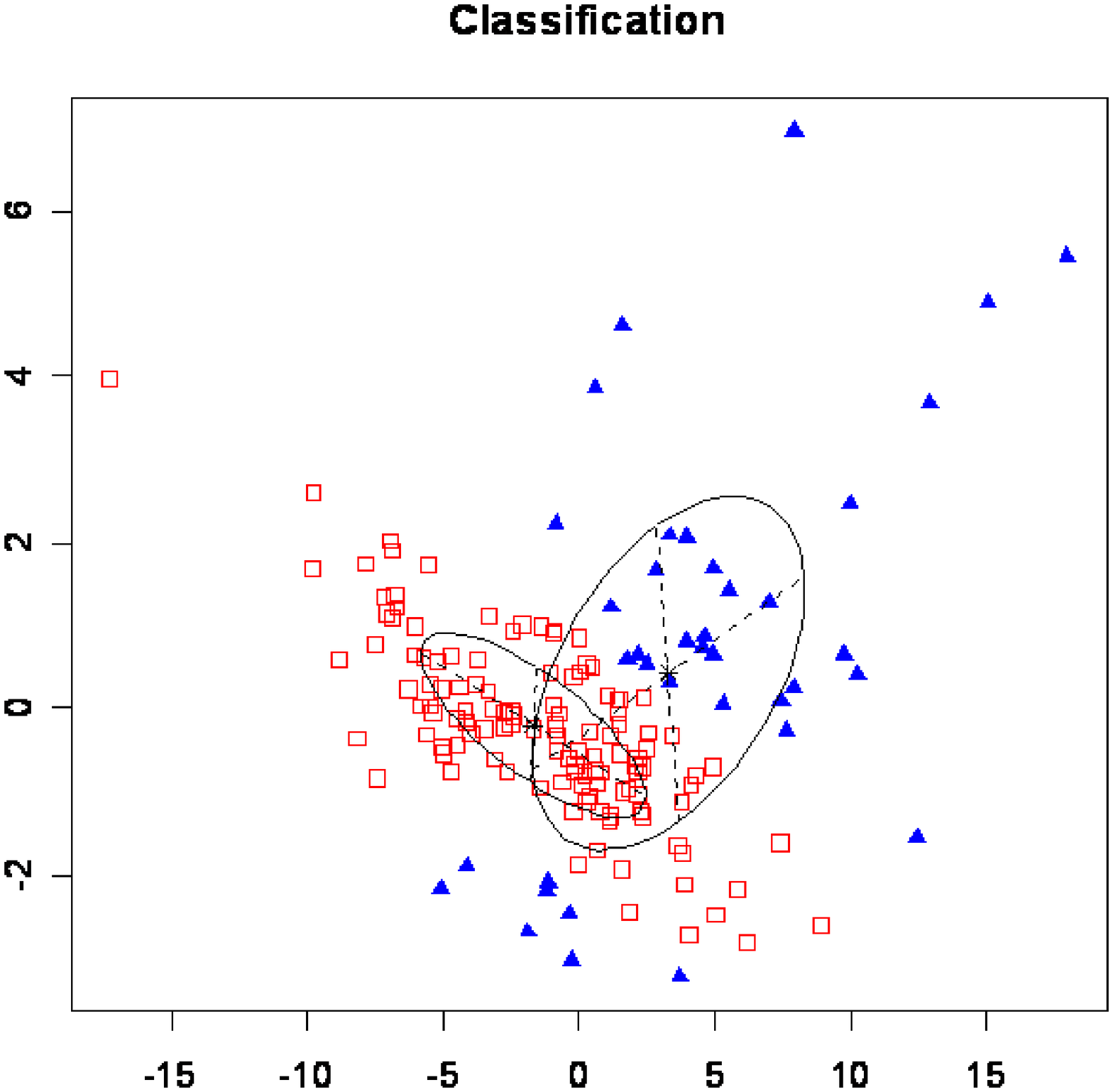}
}
\hspace{1cm}
\subfigure[GO purity of Clusters from fPCA.] % caption for subfigure b
{
    \label{fig:fpca_GO}
    \includegraphics[height=4.6cm,width=6.6cm]{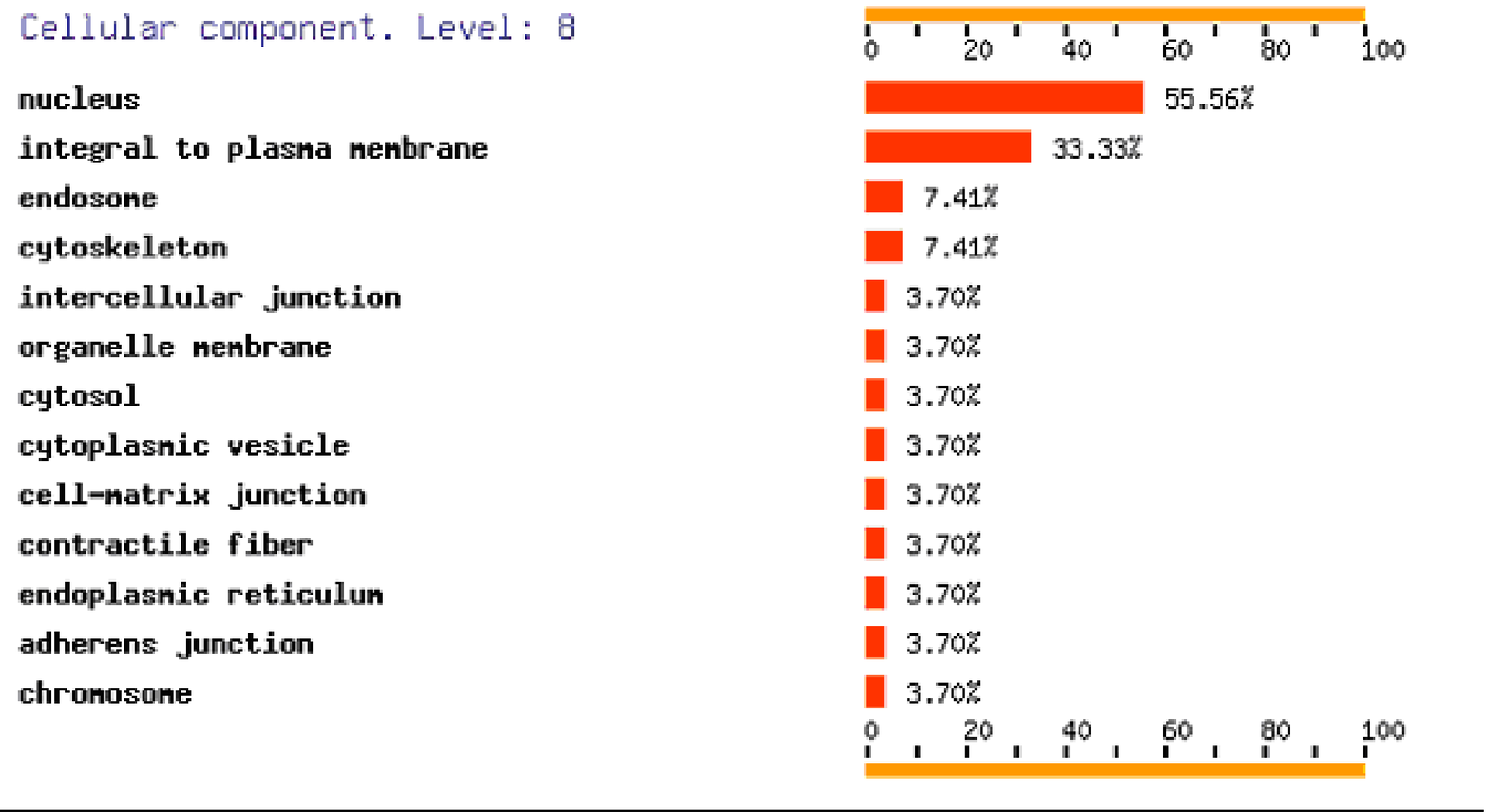}
}
\caption{BIC based clustering for fPCA embedded data and GO cluster purity.}
\label{fig:sub} % caption for the whole figure
\end{figure}

Based on the Bayesian Information criterion (BIC), we select two clusters (Fig. \ref{fig:fpca_tub_bic}) with variable shape, volume and orientation (VVV). Additionally, to ascertain the purity of clustering, we examine if co-clustered genes are in the same cellular location.  This is done using the FATIGO$+$ tool at \emph{http://babelomics.bioinfo.cipf.es/fatigoplus/}. The results of this analysis is in Fig. \ref{fig:fpca_GO}. The results indicate that only about 50\% of the co-clustered genes are co-located (in the nucleus). This casts serious doubts on the analysis methods of previous papers that use clustering in PCA/fPCA space as a method to discover novel pathway components. Biologically, the cellular proximity of two genes is essential for their interaction along a pathway. Thus, unless cellular proximity can be explicitly incorporated into this framework, such clustering can potentially fail in the discovery of true pathway components [\refcite{go_clust2}, \refcite{go_clust}].

\section{COMMENTS}
To recapitulate, there is a need to find molecular signatures that predict grade of disease/ therapeutic potential. The traditional approach to find dysregulated pathways is based on
clustering genes based on expression profiles (for the corresponding disease) -- in fPCA space. This uses the hypothesis that co-clustered (co-expressed) genes are `possibly' part of the same pathway -- since genes with similar expression profiles all belong to same pathway, i.e. their relationship is so tight that they will behave in the same coordinated manner.
A lot of literature using this hypothesis is available.% : Incorrect!!

However, this approach is questionable because in any biological process, several pathways are involved, and there are cross-interactions (cross-talk) among pathways.  Clusters can thus consist of putatively interacting pathways, not just one pathway.
Because different conditions/diseases are due to different aberrant pathways, clusters can have different sets of genes and thus several `true' pathways. This leads to incorrect inference of the biology, because every study can find  a`new' pathway based on which condition they study. In reality, there are only a few standard pathways -- however their interactions are different in different diseases/conditions.

Thus, we would need to incorporate some other prior knowledge to aid the clustering approach in achieving biological realism. One such way would be to use location information in conjunction with expression and cluster genes in that combined space -- this follows from fact that if genes have to have coordinated pathway activity, they should be nearby in cellular location.

\section{Part I: BUILDING REALISM WHILE CLUSTERING}\label{biological_realism}
As suggested in the previous sections, it would be useful to have a ``space" which respects physical cellular proximities in addition to expression similarities. This can be enabled by considering a set of annotations that describe the ``cellular location" information for each of the genes (in the macrophage activation program). One set of annotations that is well researched by the bioinformatics community is the Gene Ontology (GO) descriptors (\emph{http://www.geneontology.org/}). This is a controlled hierarchical vocabulary that annotates genes  in various  organisms by cellular component (CC), molecular function (MF) and biological process (BP), based on literature reports.

The next section examines the generation of a ``semantic similarity matrix" between genes based on their GO (CC) descriptors, to quantify the cellular proximity among them. Just like lexical word ontologies for spoken languages (e.g. WordNet at \emph{http://wordnet.princeton.edu/}), this structure imposes a tree structure on the various GO terms, thereby expressing the similarity between any two terms in the ontology as a function of their parents in the ontology tree.

The next step involves the use of manifold embedding techniques that can integrate such GO similarity along with expression-level similarity to construct an embedding of the genes as points in some space. One such technique is Laplacian Eigenmaps [\refcite{Laplacian}], also profiled in Section: \ref{Lapl_Eig} that approximate both these relationships (semantic and expression similarities). % in a mean square sense.
This is a generalization of the principal component approach in that the distance measures on such manifolds are not necessarily euclidian. After embedding the genes onto this manifold, we will then re-examine the model based clustering approach and assess the GO purity (as in section: \ref{fpca_cluster}) of the obtained clusters.

We remind the reader that the main goal here is to embed genes based on their expression profiles, but additionally weighted based on their cellular proximity -- this would be more biologically relevant for the discovery of true pathway activity. We believe that such an approach is consistent with the rationale of using integrative genomics or principled heterogeneous data integration for stronger hypothesis generation [\refcite{data_integration}].

\subsection{GO Semantic Similarity}\label{GOsim}
To quantify the notion of similarity of terms along an ontology, we appeal to a vast amount of literature that addresses such questions [\refcite{Semantic_Similarity}]. The semantic similarity of any two GO terms along the ontology hierarchy is based on the number of shared parents and the information content of  the individual GO terms (measures: Jiang Conrath, Resnik etc.). Based on the literature, we use the Jiang-Conrath  similarity measure, given by,\\

$W_{i,j} = sim(c_i,c_j) = \frac{1}{jc_{dist}(c_i,c_j)}$, with $jc_{dist}(c_i,c_j) = 2 log(p(lso(c_i,c_j))) - [log(p(c_i))+log(p(c_j))]$\\

where $c_i$ and $c_j$ are two terms (nodes) along the GO ontology tree ($i,j \in \{1,2,\ldots,168\}$). $lso(c_i,c_j)$ refers to the the information content of the last common parent of these two nodes. The information content is computed based on the probabilities of observing the individual nodes and their last common ancestor in an overall corpus.

For the $168$ genes profiled in this study, we use the R package ``GOSim" to obtain the semantic similarity matrix (size $168 \times 168$) based on GO location annotation. This similarity matrix is used to obtain the weight matrix $W$ during the Laplacian Eigenmap embedding procedure [\refcite{Laplacian}] below.

\subsection{LLE (Laplacian Eigenmaps)}\label{Lapl_Eig}
\begin{figure}[H]%embedding and BIC cirve
\centering
\subfigure[Eigenmap embedded data.] % caption for subfigure a
{
    \label{fig:LLE_embed}
    \includegraphics[width=6cm]{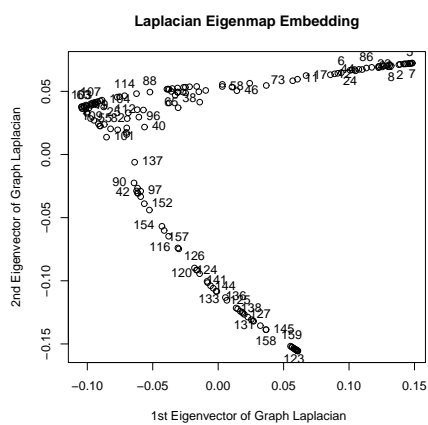}
}
\hspace{1cm}
\subfigure[BIC plot for Eigenmap based clustering.] % caption for subfigure b
{
    \label{fig:LLE_BIC}
    \includegraphics[width=6cm]{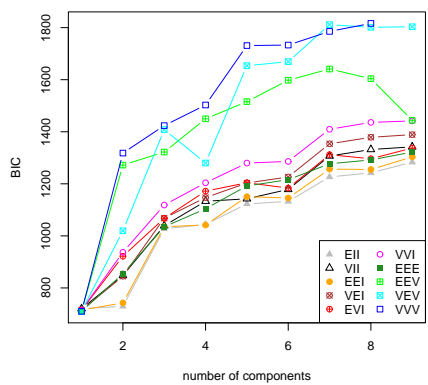}
}
\caption{Eigenmap based embedding and BIC plot.}
\label{fig:sub} % caption for the whole figure
\end{figure}

%Suppose we are investigating the role of $(K-1)$ genes in relation
%to our target gene (\textit{Gata3}) - we proceed as follows:
\begin{itemize}
%\item
%Standardize these $K$ gene expression profiles to 0 mean and unit
%variance. Notice that the Euclidean distances become the Pearson
%correlation measure.\\%\vspace{-.05in}
\item
Build the $K \times K$, $(K=168)$ dimensional weight matrix $W$ from the Gene Ontology (``Cellular Component") terms of the genes in the dataset. This distance is the ``normalized" semantic similarity alluded to above (section \ref{GOsim}).
\item
 Assign weight $W_{i,j}$,
from (1) for each gene pair $(i,j)$, for each of the ${K \choose 2 }$
gene pairs. \textit{Note}: The higher this weight, the closer the genes are.%\vspace{-.05in}
\item
Find $n$ nearest neighbors using the euclidian distance in principal component space. The scores of the functional data along the first two principal components can be interpreted as co-ordinates in a euclidian space.
\item
Form the Graph Laplacian:
\[ L_{i,j} = \left\{ \begin{array}{ll}
d_i = \sum_{k}W_{i,k} & \mbox{if $i=j$};\\
-W_{i,j} & \mbox{if $i \text{ is connected to } j$};\\
0 & \mbox{otherwise}.\end{array} \right. \]\\
%\vspace{-.05in}
\item
   Solve:
   $min_y  y^{T}Ly = \frac{1}{2}\sum_{i,j}(y_i-y_j)^2 W_{i,j} \quad
   (2)$, \\ subject to:
   \begin{itemize}
   \item
   \quad \quad $y^TDy = 1$, and \\
   \item
   \quad \quad $y^T D \textbf{1}=0$,
   \end{itemize}
   where $D_{i,i} = \sum_{j} W_{j,i}$, a diagonal weight matrix.
  % \vspace{-.05in}
\item
Embed the co-ordinates to a lower dimensional manifold, using the
solution (the Laplacian Eigenmap) obtained from the minimization
above.\\%\vspace{-.05in}
\begin{itemize}
\item
The solution to (2) is given by the \textit{d} generalized
eigenvectors associated with the \textit{d} smallest generalized
eigenvalues solving $L \textbf{y} =\lambda D \textbf{y}$ (neglecting the zero eigenvalue and its eigenvector).
\item
If  $\textbf{y}=[y_1,\ldots,y_d]$ is the collection of these
eigenvectors, then the embedding is given by : \\$y_i =
(y_{i1},\ldots,y_{id})^T$, i.e., the $d$ dimensional representation
of the $i^{th}$ data point (gene).
\end{itemize}
\item
In our representation, we take dimensionality, $d$ = 2 and number of
neighbors, $n$ = 5. The final embedding of the functional data based on expression and location modalities is shown in Fig. \ref{fig:LLE_embed}.
\end{itemize}

\subsection{Results with Laplacian Eigenmaps}

After transforming the original scores (from fPCA) based on the GO semantic similarity matrix $W$, the $2D$ representation of the gene data is shown in Fig. \ref{fig:LLE_embed}. Based on this new embedding, which leads to a very different visualization of the data compared to Fig. \ref{fig:score_fpca}, we can once again use model based clustering in the new space.

%Clustering, GO purity

\begin{figure}[H]%clusters and GO purity
\centering
\subfigure[Clustering on Eigenmap embedded data.] % caption for subfigure a
{
    \label{fig:LLE_cluster}
    \includegraphics[width=6cm]{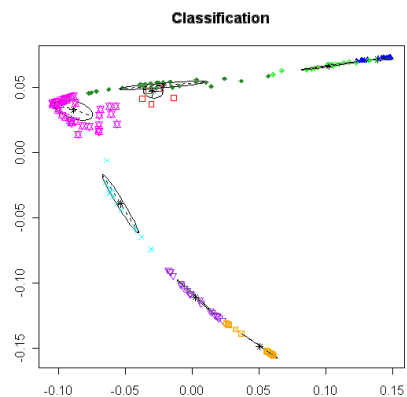}
}
\hspace{1cm}
\subfigure[GO purity of Eigenmap based clustering.] % caption for subfigure b
{
    \label{fig:LLE_cluster_GO}
    \includegraphics[height=4.2cm,width=8cm]{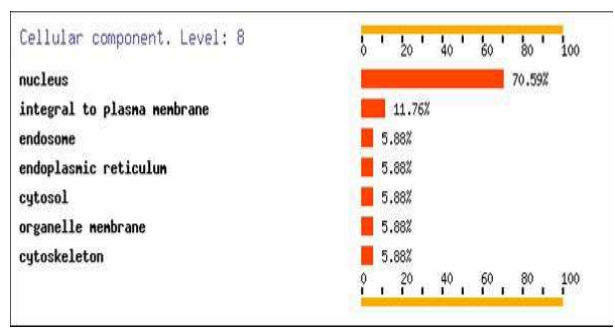}
}
\caption{BIC based clustering for Eigenmap embedded data  and GO purity.}
\label{fig:sub} % caption for the whole figure
\end{figure}

Based on this, the BIC plot for the Eigenmap embedded data is shown in Fig. \ref{fig:LLE_BIC}. The highest BIC corresponds to $8$ clusters (Fig. \ref{fig:LLE_cluster}) with variable shape, volume and orientation (VVV). An examination of the GO (``cellular component" annotation) is shown in Fig. \ref{fig:LLE_cluster_GO}. This indicates that the GO enrichment of genes that are close by (like in nucleus) is much higher ($\sim 70\%$ compared to $55\%$ before). Furthermore, interrogation of a cluster in the new assignment (after Eigenmap embedding) clearly identifies known components of the cytokine pathway [\refcite{Huang_Hacohen}] (several interleukin members). This shows that an embedding, that respects location as well as expression simultaneously, identifies closely interacting pathway components via clustering. This aids in the development of biologically relevant hypotheses.

%As can be seen,
%explain mclust results and the purity of each cluster, and hat GO purity is much higher, so more biologically plausible pathway discovery.

\section{Part II: QUERYING PATHWAY ACTIVITY}\label{part2}

The first part of this work presents a principled framework to embed gene relationships based on expression and cellular location. This framework can now be extended to understand the co-ordinated activity of genes constituting a pathway. We note that this question has not hitherto been asked in this context previously. Previous methods have only looked at identifying pathway genes based on expression similarities. In the light of the previous analysis suggesting that plain clustering in expression space without regard to physical proximity might not be biologically relevant, this framework enables the integration of multiple modalities to obtain much more relevant results. Embedding the data using a principled approach enables the formulation of more complex queries such as coordinated pathway activity.

The key question that can now be addressed under the above framework is: How strongly is pathway `X' dysregulated between normal cells and diseased (tuberculosis infected) cells. A pathway is a set of interacting genes. More generally, one can query for the co-ordinated activity of any subset of interesting genes (generalizing the approach of [\refcite{GSEA}, \refcite{GSA}].

\subsection{Querying Pathway Activity}
In order to query a pathway's activity, we can first obtain the known components (genes) of the pathway using resources such as the KEGG (\emph{http://www.genome.ad.jp/kegg/pathway.html}) or BioCarta (\emph{http://cgap.nci.nih.gov/Pathways/BioCarta\_Pathways}) pathway repositories.

For any query pathway \emph{P} (consisting of $k$ genes),we can find the co-ordinates of this subset on the manifold embedding obtained above. Let $C_p$ be the inter-point distance matrix ($k \times k$) of the $k$-gene pathway in the eigenmap for the control condition. Let $T_p$ be the inter-point distance matrix of the k-gene pathway under the perturbation (tuberculosis infection).

In this setting, the question of querying pathway activity translates to the following question:
How are the gene-gene associations among these $k$-components of the pathway ``different" between control ($C_p$) and case ($T_p$).

This is addressed in the following section. The idea is to find a metric of similarity (or distance) between the distance matrices ($C_p$ and $T_p$). The Mantel correlation test has been used in ecological studies as a similarity metric. The other is a hitherto unused metric, termed the \textit{logDet} divergence, that has been used in other applications (to quantify the distributional divergence between two probability distributions).

\subsection{Quantifying difference in association matrices}

Based on the above, the Mantel's test is used in ecological analysis (R package ``vegan") to compare two (or more) spatial proximity matrices. In our context, we are interested to ask if the gene set that is close in one space (control) is also close in the other space (tuberculosis).

The Mantel correlation coefficient between two ($k \times k$) matrices $X$ and $Y$ -  $`r'$ is given by [\refcite{Mantel_test} ]:

$r = \frac{1}{(n-1)}\sum_{i < j} \frac{x_{ij}-\bar{x}}{s_x}. \frac{y_{ij}-\bar{y}}{s_y}$

Here, $X = C_p$ and $Y = T_p$. $s_x$ and $s_y$  are the standard deviations from the entries of the matrices $X$ and $Y$, respectively (for normalization); $n = \frac{k(k-1)}{2}$. The higher the value of $`r'$, the more similar the two distance matrices are.\\

%Divergence of normal distributions parameterized by correlation matrices: logDet divergence

%Can also find the amount of divergence as well as its significance via bootstrapping.
Finally, we can estimate a significance of $r(T_p,C_p)$ via bootstrapping. This would involve the following steps:
\begin{itemize}
\item
Repeat the following procedure \emph{B}$(=1000)$ times (with index $b = 1,\ldots,B$):
\begin{itemize}
%\item
%sample a permutation $z^b$ from a uniform distribution over $\mathbb{Z_N}$
\item
Generate resampled (with replacement) versions of the matrices $C_p$, $T_p$, denoted by $C_p^b$, $T_{p}^b$ respectively.
\item
Compute the statistic $\theta^b$ = $r(T_p^b,C_p^b)$.
%\item
%compute the statistic $\theta^b$ = $\hat{I}_B(X^N \rightarrow Y_{z^m}^N)$ OR $\hat{I}_B(X_{z_1^m}^N \rightarrow Y_{z_2^m}^N)$. %OR $\hat{I}_B(X_{z^m}^N \rightarrow Y_{z^m}^N)$
\end{itemize}
\item
Construct an empirical CDF (cumulative distribution function) from these bootstrapped sample statistics, as
$F_\Theta(\theta) = P(\Theta \le \theta) = \frac{1}{B}\sum_{b=1}^B I_{x \ge 0}(x = \theta-\theta^b)$, where $I$ is an indicator random variable on its argument $x$.
\item
Compute the true detection statistic (on the original time series) $\theta_0 = r(T_p,C_p)$ and its corresponding $p$-value ($p_0=1-F_\Theta(\theta_0)) $ under the empirical null distribution $F_\Theta(\theta)$.
\item
If $F_\Theta(\theta_0)  \ge (1-\alpha)$, then we have that the true mantel correlation value is significant at level $\alpha$, leading to rejection of null-hypothesis (no association).
\end{itemize}

\subsection{logDet divergence}

The \textit{logDet} divergence has recently received a lot of interest in the machine learning community, mainly with regard to metric learning problems [\refcite{metric_learning}]. To see an example of how they arise, consider the the Kullback-Liebler (KL) divergence between two multivariate Gaussian densities $p(x; C_p)$ and $p(x; T_p)$. This is given by:\\ $KL(p(x; C_p)||p(x; T_p))=\frac{1}{2}D_{ld}(C_p,T_p)$, where,

$D_{ld}$  is the \textit{logDet} divergence between two positive definite matrices $C_p$ and $T_p$ defined by:

$D_{ld}(C_p,T_p) = tr(C_pT_p^{-1})-log det(C_pT_p^{-1}) – k$
\quad ($k$ is the rank of the matrices $C_p$ and $T_p$)

We note that the distance matrices $C_p$ and $T_p$ are positive semi-definite, but can be made positive definite via the addition of a constant term to its non-diagonal elements [\refcite{Cailiez_method}] (Cailiez method). Before using the \textit{logDet} criterion, the distance matrices need to be converted into correlation matrices (this can be done via a scaled exponential transformation [\refcite{Laplacian}]. Additionally, since the KL divergence is not intrinsically symmetric, the symmetrized version, $LD_{dist}(T_p,C_p) = \frac{1}{2}D_{ld}(C_p,T_p)+\frac{1}{2}D_{ld}(T_p,C_p)$ can be used instead. The higher the value of this symmetrized distance, the higher the dissimilarity between $C_p$ and $T_p$.\\

Finally, we can estimate a significance of $LD_{dist}(T_p,C_p)$ via bootstrapping. This would involve the following steps:
\begin{itemize}
\item
Repeat the following procedure \emph{B}$(=1000)$ times (with index $b = 1,\ldots,B$):
\begin{enumerate}
%\item
%sample a permutation $z^b$ from a uniform distribution over $\mathbb{Z_N}$
\item
Generate resampled (with replacement) versions of the matrices $C_p$, $T_p$, denoted by $C_p^b$, $T_{p}^b$ respectively.
\item
Compute the statistic $\theta^b$ = $LD_{dist}(T_p^b,C_p^b)$.
%\item
%compute the statistic $\theta^b$ = $\hat{I}_B(X^N \rightarrow Y_{z^m}^N)$ OR $\hat{I}_B(X_{z_1^m}^N \rightarrow Y_{z_2^m}^N)$. %OR $\hat{I}_B(X_{z^m}^N \rightarrow Y_{z^m}^N)$
\end{enumerate}
\item
Construct an empirical CDF (cumulative distribution function) from these bootstrapped sample statistics, as
$F_\Theta(\theta) = P(\Theta \le \theta) = \frac{1}{B}\sum_{b=1}^B I_{x \ge 0}(x = \theta-\theta^b)$, where $I$ is an indicator random variable on its argument $x$.
\item
Compute the true detection statistic (on the original time series) $\theta_0 = LD_{dist}(T_p,C_p)$ and its corresponding $p$-value ($p_0=1-F_\Theta(\theta_0)) $ under the empirical null distribution $F_\Theta(\theta)$.
\item
If $F_\Theta(\theta_0)  \ge (1-\alpha)$, then we have that the true \textit{logDet} value is significant at level $\alpha$, leading to rejection of null-hypothesis (complete association).
\end{itemize}

%Estimate Mantel correlation and \textit{logDet} divergence significance via bootstrapping.

\section{CASE STUDY:TLR PATHWAY}

%give Biocarta pictures of TLR and Mapk

\begin{figure}[H]%give TLR and Mapk pathway pics
\centering
\subfigure[TLR pathway (\copyright Biocarta).] % caption for subfigure a
{
    \label{fig:tlr_biocarta}
    \includegraphics[width=6.6cm]{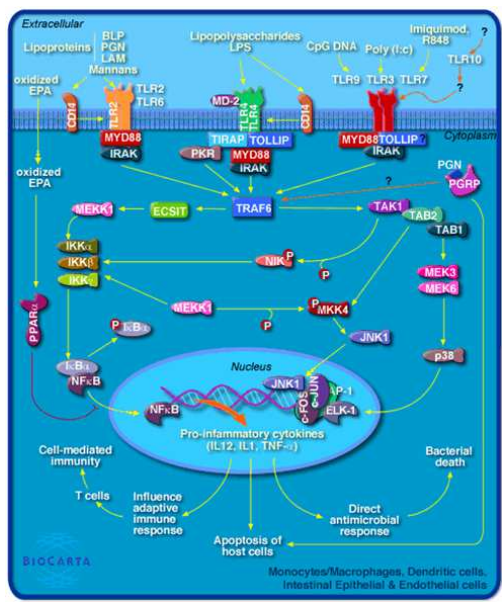}
}
\hspace{1cm}
%\subfigure[Mapk pathway (Biocarta).] % caption for subfigure b
%{
%    \label{fig:sub:b}
%    \includegraphics[width=8cm]{mapk_pathway.eps}
%}
\subfigure[Null Histogram of the \textit{logDet} divergence for the TLR pathway. True value=0.5513.
Is Activated Early in Immune Response] % caption for subfigure a
{
    \label{fig:tlr_hist}
    \includegraphics[width=6cm]{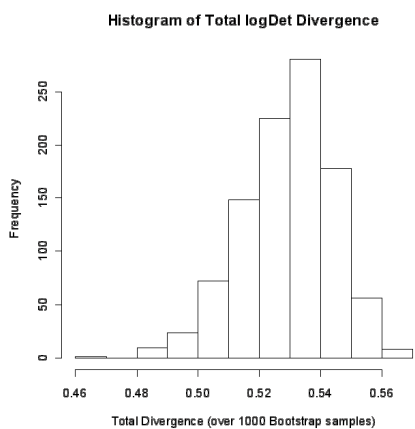}
}
\caption{The TLR pathway and its Divergence between normal and case conditions.}
\label{fig:sub} % caption for the whole figure
\end{figure}

As an example of querying pathway activity, we analyse the toll-like receptor (TLR) pathway.
Members of the toll-like receptor (TLR) gene family convey signals stimulated by various pathogenic factors, activating signal transduction pathways that result in transcriptional regulation and stimulate innate immune function (Fig. \ref{fig:tlr_biocarta}). Hence it is one of the earliest and most strongly activated pathways during immune response. %Explain TLR in some level.

For this pathway, we get the subset of genes that are common between the Biocarta  TLR catalog and the set of $168$ genes. For those $7$ genes, we can find the inter-point distance matrices along the normal and infection cases. The value (also referred to as the \emph{true value}) of the mantel correlation and \textit{logDet} divergence (between the normal and control states) is $0.0447$ and $0.5513$ respectively, suggesting that the association between these two distance matrices is fairly low. Additionally, these correlation and divergence values are significant at the $0.05$ level with respect to the null distribution (Fig. \ref{fig:tlr_hist}). Hence, these two measures are useful at picking up a truly activated pathway between these two conditions. In the same way, one can obtain these two measures for any chosen pathway of interest, yielding results that are concordant with literature [\refcite{Huang_Hacohen}]. This is shown in Table $I$.% \ref{tbl3}.

%
%Bootstrapping results from Toll-like Receptor (TLR) and Mapk pathways

\begin{figure}[H]%bootstrap results
\centering
\subfigure[Null Histogram for the TLR pathway components. True value=0.5513.
Is Activated Early in IR.
] % caption for subfigure a
{
    \label{fig:sub:a}
    \includegraphics[scale=0.2]{tlr_hist.eps}
}
\hspace{1cm}
\subfigure[Null Histogram for the Mapk pathway components. True-value: 0.018.
Not activated early.] % caption for subfigure b
{
    \label{fig:sub:b}
    \includegraphics[scale=0.2]{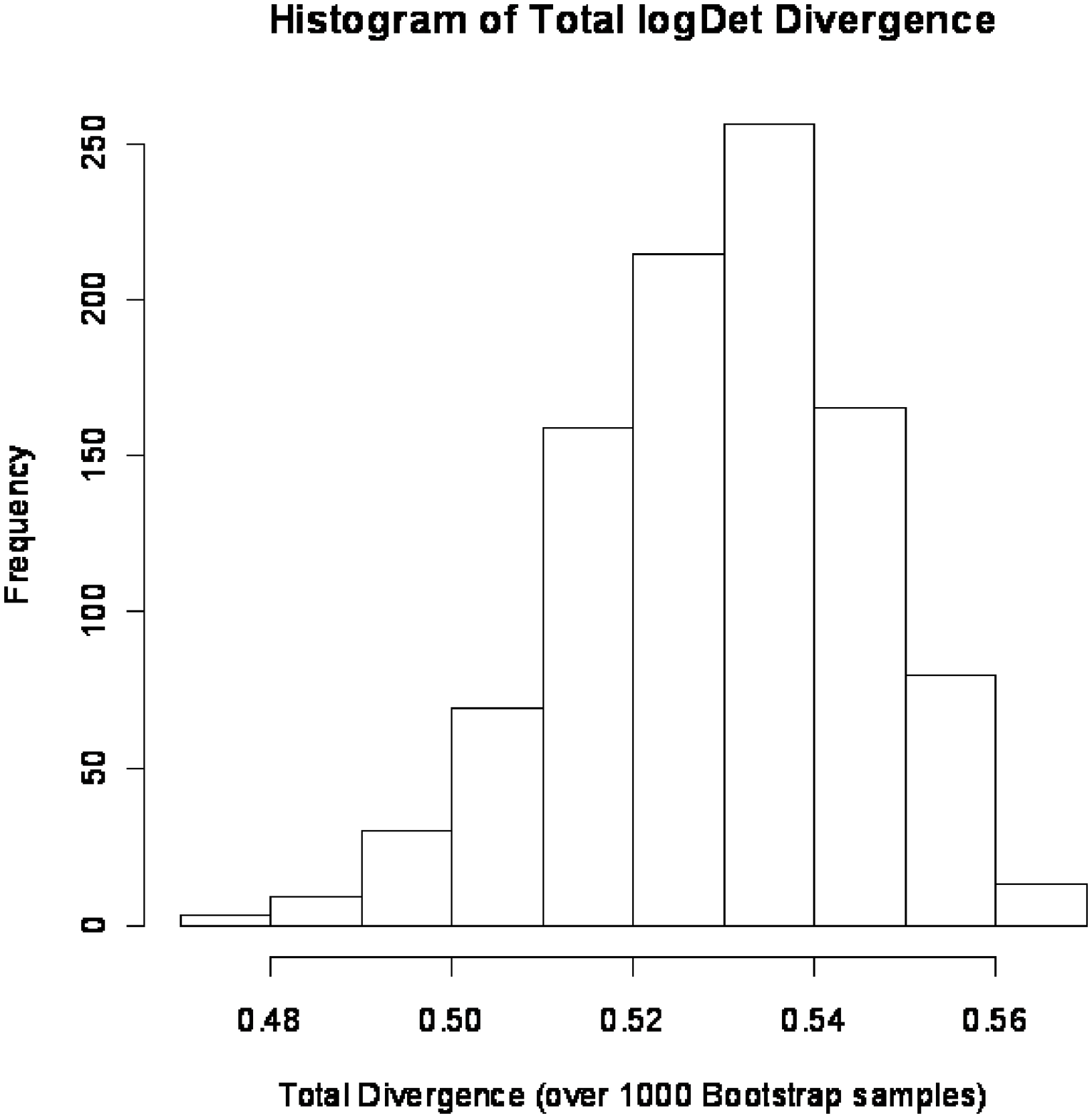}
}
\caption{Bootstrapping results from Toll-like Receptor (TLR) and Mapk pathways.}
\label{fig:sub} % caption for the whole figure
\end{figure}

%Rank pathways based on Divergence (GIVE TABLE)

\begin{table*}%[H]%[!t!b]
\tbl{Mantel and \textit{logDet} values of some interesting pathways.}
{\begin{tabular}{ccc} \hline
   Pathway Name      &     Mantel test (value and significance)    & \textit{logDet} divergence (value and significance) \\
   \hline
   \textit{Apoptosis}                       & 0.5961(0.038)     & 0.149(0.025)\\
   \textit{Toll-like receptor (TLR)}        &    0.047(0.0255)  & 0.5513(0.032)\\                         \textit{Mapk}                            &  0.3373(0.142)    & 0.018(0.010)\\                               \textit{T/B-cell}                        &  0.271(0.071)     & 0.1523(0.006)\\
   \hline %have more entries in this t able
\end{tabular}} \label{tbl3}%\\ \\ \\ \\
\label{Tab:gata3_TF}
\end{table*}

\section{CONCLUSIONS}\label{conclusions}
In Part $I$ of this work, we demonstrated a generalizable method to infer pathway components (or cross-talking pathways). Using Laplacian Eigenmaps, we were able to co-embed genes  based on expression and location modalities. Model based clustering on embedded data further confirms that genes that are co-clustered also have higher purity with respect to cellular location. Additionally, some of the co-clustered genes belong to canonical pathways.
%Higher cluster purity when expression data is integrated with cellular location information.

From the $2D$ space obtained in Part $I$, we develop a novel framework (Part $II$) to query the activity of any gene set (or pathway of interest) across biological conditions using the Mantel correlation and \textit{logDet} divergence.
%Part II:
%A generalizable framework to investigate the behavior of any 'canonical' pathway in interesting biological conditions.

The overall contribution of this work is the development of a complete workflow that combines functional data analysis on expression data with ontology to yield biologically relevant results via heterogeneous data integration. Though there has been some previous work [\refcite{go_clust}, \refcite{go_clust2}] combining gene expression with ontology to understand gene co-regulation, we are aiming to do this for whole pathways or gene sets.
%Novel Contribution: A complete workflow from expression profiles, FDA, GO ontology to pathway ranking shown.

\section{EXTENSIONS AND FUTURE WORK}
The methods developed in this work, both for embedding genes based on expression and cellular location are applicable for any study of interest and thus easily generalizable. Additionally, the query for the co-ordinated activity of any gene set of interest (pathway or otherwise) is also generalizable since it only examines the association of the inter-gene distance matrix (along the manifold) between case and condition. This procedure also enables a relative ranking of multiple pathways, thereby allowing for simultaneous queries.

This work also expands on previous approaches in heterogeneous data integration, combining modalities like gene ontology with gene expression specifically for pathway query along an biological process of interest. This could further enable efforts to understand pathogenesis through the modulation of pathway activity between the normal and diseased cell state.

%Can rank canonical pathways dysregulated between different conditions.

Finally, though this work uses the Mantel correlation and the \textit{logDet} divergence for determining the difference in distance matrices, several other methods such as Procrustes alignment [\refcite{procrustes}], or the Jensen-Shannon distributional divergence can be used for the same purpose. It would be interesting to see if any of the methods make fewer distributional assumptions on the structure of the inter-point distance matrices.

%Other Metrics: Procrustes, Jensen-Shannon Divergence

\section*{AVAILABILITY}
The source code of the analysis tools (in R $2.6$) is available on request. The gene expression data is publicly available at: \\ \emph{http://web.wi.mit.edu/young/}.

\section*{ACKNOWLEDGEMENTS}
The author would like to thank Prof. Tailen Hsing for his offering of the STAT $700$ course at the University of Michigan, as well as for his comments and feedback during the project. %Feedback from several classmates is also gratefully acknowledged.
The author's support as a Lane Fellow in Computational Biology at Carnegie Mellon University is gratefully acknowledged. We also thank Prof. Robert Murphy for feedback on an earlier draft of this manuscript.

%%%%%%%%%%%%%%%%%%%%%%%%%%%%%%%%%%%%%%%%%%%%%%%%%%%%%%%%%%%%%%%%%%%%%%%%%%%%%%%%%%%%%
%
%     please remove the " % " symbol from \centerline{\includegraphics{fig01.eps}}
%     as it may ignore the figures.
%
%%%%%%%%%%%%%%%%%%%%%%%%%%%%%%%%%%%%%%%%%%%%%%%%%%%%%%%%%%%%%%%%%%%%%%%%%%%%%%%%%%%%%%

%\section{REFERENCES}
%Authors must pay particular attention to the accuracy of references
%which should be checked before final submission of the finished
%manuscript. The following are some examples of references.
%
%\subsubsection*{Journal Article:}
%Rogers WD, Watson HK. Radical styloid impingement
%after triscaphe arthrodesis. {\it J Hand Surg} 1989; 14A: 297--301.
%
%\subsubsection*{Books:}
%Taleisnik J. {\it The Wrist.} Churchill Livingstone, New York. 1985:
%15--20.
%
%\subsubsection*{Chapters in Edited Book:}
%Beckenbough RD, Linscheid RL.
%Arthroplasty in the hand and wrist. In: Green DP (ed.), Operative
%Hand Surgery, 2nd ed. Churchill Livingstone, New York. 1988:
%167--214.
%
%\noindent Washburn SL. Longevity in primates. In: Mc Gaugh JL, Klesler SB
%(eds.), Aging: Biology and Behavior. Academic Press, New York. 1981:
%11--29.

\end{multicols}
\end{document}